# Phonon-Bottleneck-Governed Ultrafast Hot-Carrier Super-Diffusion in Transition Metal Dichalcogenides

Yuwei Zhang[1, #], Dongyang Wan[1, #, *], Tao Zhou[1], Hao Wu[1], Junpeng Lu[1, 2, *], Zhenhua Ni[1,2, *]

[1] Key Laboratory of Quantum Materials and Devices of Ministry of Education, School of Physics, Southeast University, Nanjing 211189, China

[2] School of Electronic Science and Engineering, Southeast University, Nanjing 210096, China

[#]These authors contributed equally: Yuwei Zhang, Dongyang Wan

[*]e-mail: wandy@seu.edu.cn, phyljp@seu.edu.cn, zhni@seu.edu.cn.

**Abstract**

Two-dimensional transition metal dichalcogenides (TMDCs) are promising for low-power optoelectronics, yet their operational speed is widely considered constrained by low room-temperature mobilities and carrier transit delays. Here, by combining on-chip terahertz optoelectronic sampling with thermally evaporated Ohmic contacts, we eliminate external parasitic delays and directly capture the intrinsic interfacial photoresponse in unencapsulated TMDCs under zero bias. The devices achieve ultrafast relaxation lifetimes of 48.5 ps in $MoS_2$/Au and 14.2 ps in $MoSe_2$/Ag, translating to intrinsic 3-dB bandwidths of 4.4 GHz and 7.5 GHz, respectively. Spatial scanning and bias-dependent measurements show that this response is position-independent and bias-immune, ruling out conventional drift-limited transport and identifying hot-carrier super-diffusion driven by an interfacial electron temperature gradient as the operative mechanism. Furthermore, ultrafast pump-probe spectroscopy reveals that the macroscopic response time is quantitatively synchronized with the microscopic optical-to-acoustic phonon scattering lifetime governed by the intrinsic phonon bottleneck. Our findings establish phonon engineering as a viable paradigm to tailor non-equilibrium optoelectronic dynamics, offering a blueprint for zero-bias, ultrafast, self-powered devices.

## Introduction

Two-dimensional (2D) transition metal dichalcogenides (TMDCs) are widely recognized as ideal platforms for next-generation low-power, self-powered optoelectronic devices, owing to their layer-dependent electronic band structures, pronounced light-matter interactions, and intrinsic dark currents that are orders of magnitude lower than those of zero-bandgap semimetals [1,2]. Compared with early zero-bandgap 2D materials such as graphene, the well-defined bandgaps of TMDCs effectively suppress dark currents and yield superior optoelectronic on/off ratios [3,4], offering distinct advantages for passive, integrated optoelectronics. However, governed by strong carrier-phonon scattering and relatively large carrier effective masses, the intrinsic room-temperature band-edge drift mobility of TMDCs is modest (typically 10-100 $cm^2V^{-1}s^{-1}$) [5,6], which has long imposed a fundamental bottleneck on their operational speed. In conventional photovoltaic (PV) and photoconductive (PC) devices that rely on the directional drift of band-edge carriers, carrier channel-transit times and the slow trapping/de-trapping dynamics associated with interfacial defect states limit the response times of typical TMDC photodetectors to the microsecond or even millisecond regime [7,8], falling far short of the requirements for high-speed optical interconnects and ultrafast information processing.

To break through this speed limit, extensive efforts have been devoted to investigating the interfacial photoresponse mechanisms and transient dynamics in 2D TMDCs. Using scanning photocurrent microscopy coupled with electrostatic gating, prior studies have identified the coexistence of photothermoelectric (PTE) and photovoltaic (PV) mechanisms at contact edges under steady-state illumination [9,10], highlighting the gate tunability of thermoelectric transport [11]; low-frequency electrical transient measurements have further elucidated trap-assisted recombination and prolonged photoconductive behaviors induced by interfacial defect states [12]. On the other hand, while ultrafast optical pump-probe spectroscopy has revealed non-equilibrium hot-carrier energy relaxation and super-diffusive transport within TMDCs [13,14], purely optical techniques cannot probe the actual electrical extraction of carriers across metal-semiconductor interfaces [15]. Constrained by conventional electrical characterization paradigms, whether photoexcited hot carriers can be extracted prior to complete energy dissipation, what microscopic mechanisms dictate their interfacial extraction dynamics, and whether this process can genuinely overcome the speed limits imposed by low band-edge mobility have remained elusive due to the lack of direct time-domain experimental interrogation.

Directly resolving these ultrafast dynamics faces dual barriers from both physical and technological perspectives. Physically, conventional metal contacts inevitably form Schottky barriers and built-in electric fields, inducing strong drift-driven photovoltaic currents that mask non-equilibrium diffusion signals; concurrently, depletion-region transit delays and defect trapping drag the response speed down into low-frequency domains [16]. Technologically, conventional electrical readout circuits are severely constrained by parasitic capacitances and external *RC* time constants, whose nanosecond-scale temporal resolution cannot capture interface transport operating at picosecond or sub-picosecond timescales [17,18]. Intertwined, these obstacles have rendered the non-equilibrium hot-carrier extraction dynamics at TMDC interfaces an experimental blind spot.

Here, by synergizing on-chip terahertz time-domain optoelectronic sampling with high-quality Ohmic contact engineering, we circumvent these fundamental bottlenecks to directly resolve the intrinsic ultrafast optical-to-electrical conversion dynamics at TMDC heterojunction interfaces under zero bias [19,20]. In unencapsulated $MoS_2$/Au and $MoSe_2$/Ag configurations, the devices exhibit ultrafast decay lifetimes as short as 14.2 ps and intrinsic 3-dB bandwidths up to 7.5 GHz. Systematic spatial mapping and bias modulation measurements confirm that this response originates from the super-diffusive extraction of non-equilibrium hot carriers, fundamentally circumventing the transport limitations imposed by low band-edge drift mobilities [21,22]. Furthermore, combined with thickness-dependent ultrafast pump-probe spectroscopy, we reveal that the macroscopic electrical extraction lifetime is strictly synchronized point-by-point with the microscopic optical-to-acoustic phonon scattering lifetime, demonstrating that the intrinsic hot-phonon bottleneck within polar lattices is the pivotal physical mechanism sustaining super-diffusion and dictating the ultimate speed limit [23,24]. Consequently, this work establishes "phonon engineering" as a design paradigm for optoelectronic devices, offering explicit material and physical blueprints to transcend carrier mobility ceilings and develop next-generation self-powered, ultrafast optoelectronic integrated chips.

## Results

*On-Chip Terahertz Sampling of Ultrafast Interfacial Photocurrents*

Two-dimensional TMDCs exhibit layer-dependent band structures and vanishingly low dark currents; however, their optical-to-electrical conversion speed has long been constrained by two physical hurdles. Externally, contact resistances and parasitic capacitances introduce nanosecond-scale *RC* delays that mask the primary interfacial dynamics. Internally, the modest room-temperature band-edge drift mobility of TMDCs subjects conventional devices to transit-time limits, restricting their operational bandwidth to the sub-gigahertz regime [5].

To eliminate parasitic *RC* limitations and directly resolve the intrinsic interfacial charge transfer, we implemented an on-chip terahertz time-domain optoelectronic sampling platform coupled with a low-temperature-grown GaAs (LT-GaAs) photoconductive switch (Fig. 1a). Mechanically exfoliated, unencapsulated TMDC flakes were integrated onto 285-nm $SiO_2$/Si substrates, with source and drain electrodes deposited via vacuum thermal evaporation to establish high-quality Ohmic interfaces (Ni/Au for $MoS_2$ and Ag for $MoSe_2$) [12, 16]. The thermally evaporated contacts ensure intimate metal-semiconductor interfaces with minimal interfacial defect states, thereby eliminating built-in electric fields and parasitic Schottky barrier effects. The on-chip readout circuit incorporates a 2-μm-thick MBE-grown LT-GaAs layer, whose defect-assisted non-equilibrium carrier lifetime of <2 ps enables sub-picosecond temporal resolution [18]. A femtosecond pump beam (λ = 680 nm) was focused locally at the metal-semiconductor edge, while a synchronized, time-delayed probe beam triggered the LT-GaAs switch to sample the transient photocurrent with high signal-to-noise ratio.

Under strict zero-bias conditions ($V_{bias}$ = 0 V), the time-resolved photocurrent waveforms display an ultrafast pulse rise followed by a rapid exponential decay (Fig. 1b). Deconvolution yields characteristic relaxation lifetimes of 48.5 ps for the $MoS_2$/Au interface and 14.2 ps for $MoSe_2$/Ag. Fast Fourier transforms (FFT) of the background-subtracted transient pulses reveal intrinsic 3-dB bandwidths of 4.4 GHz and 7.5 GHz, respectively (Fig. 1c). In layered semiconductors with room-temperature mobilities below 100 $cm^2V^{-1}s^{-1}$, conventional cold-carrier diffusion requires hundreds of picoseconds to extract charges across micron-scale distances. The realization of a multi-GHz photoresponse at zero bias demonstrates that interfacial charge extraction completely circumvents band-edge drift limits and is governed by non-equilibrium hot-carrier transport [14].

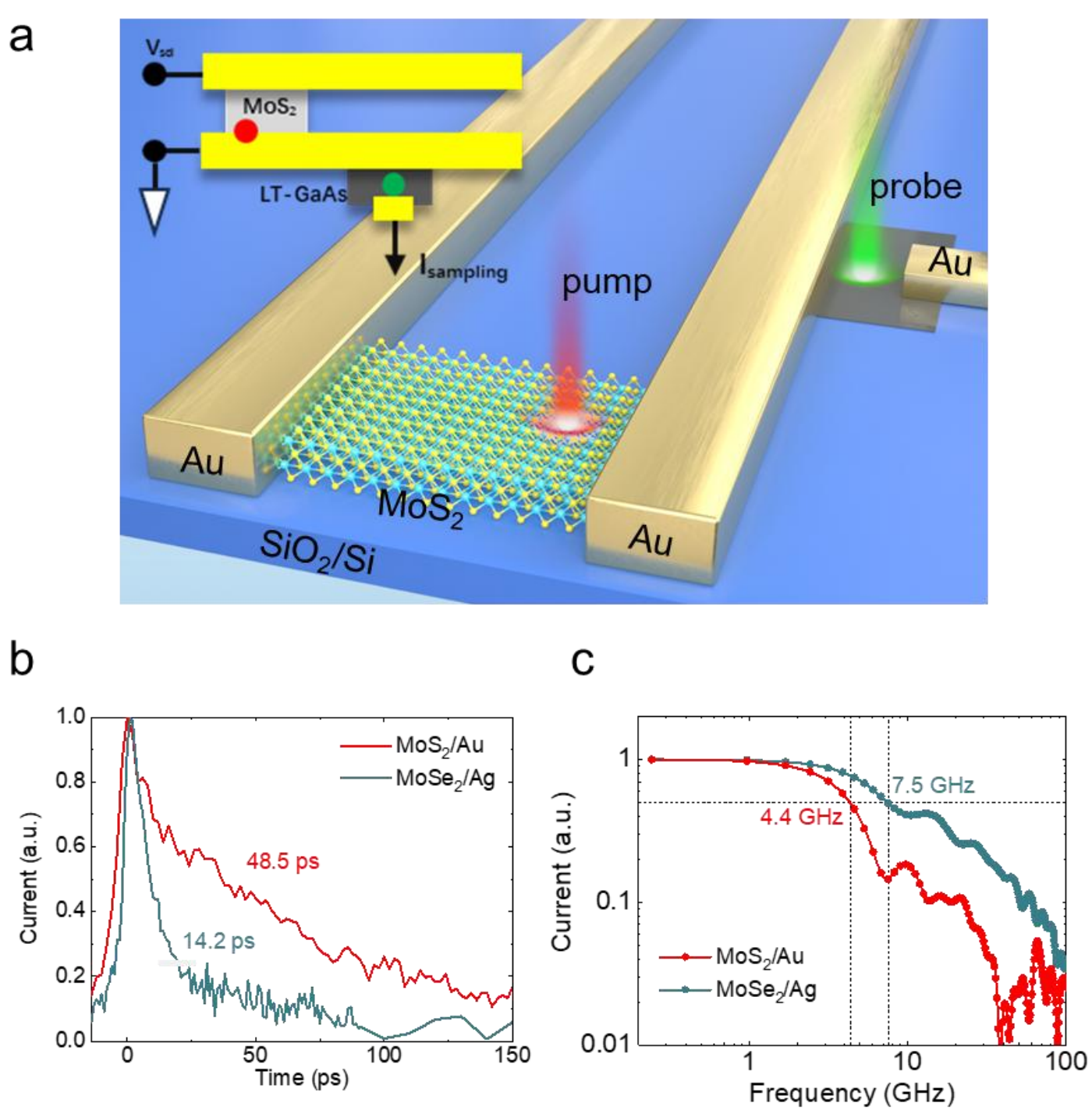

**Figure 1 | On-chip terahertz sampling of GHz-bandwidth interfacial photocurrents.**

**a.** Schematic of the device architecture. A 285-nm $SiO_2$/Si wafer serves as the substrate, with the bottom Si layer utilized for applying a bottom-gate voltage. The active device region consists of an unencapsulated $MoS_2$ flake contacted by Ni/Au (10 nm/200 nm) electrodes. As depicted in the inset, the three electrodes are connected to the bias voltage, ground, and current readout circuit, respectively. A 2-µm-thick LT-GaAs layer is integrated into the current readout circuit as a photoconductive switch. **b.** Normalized time-resolved photocurrents for the $MoS_2$/Au and $MoSe_2$/Ag devices, with the optical excitation focused precisely at the metal-semiconductor interface. **c.** Corresponding frequency-domain spectra obtained via fast Fourier transform (FFT) of the data in **b**, revealing 3-dB bandwidths of 4.4 GHz for the $MoS_2$/Au device and 7.5 GHz for the $MoSe_2$/Ag device. Experimental parameters: $V_{Bias}$ = 0 V; $V_{gate}$ = 0 V; $\lambda_{laser}$ = 680 nm; pump power: $P_{laser}$ = 1 mW.

*Experimental Validation of Hot-Carrier Super-Diffusion*

To verify whether this ultrafast response originates from photovoltaic (PV) drift driven by residual interfacial band bending, we evaluated asymmetric Schottky-contacted devices alongside our thermally evaporated Ohmic devices (Fig. 2a). In Schottky-type architectures, interfacial built-in electric fields drive directional drift separation, resulting in prolonged transient traces with nanosecond tails indicative of trap-assisted recombination [10]. In contrast, the Ohmic devices exhibit clean single-exponential decay profiles free of slow drift tails (Fig. 2b). The absence of a Schottky barrier directly verifies that the ultrafast transient pulse is decoupled from electric-field-driven drift [16].

Spatially resolved transient photocurrent measurements further substantiate the non-drift nature of the response. The pump spot was translated across the metal-semiconductor junction over a range of ±4 μm (Fig. 2c). While the peak photocurrent decreases exponentially as the excitation beam moves away from the junction into the TMDC channel, the extracted relaxation lifetime remains virtually constant at ~44.6 ps across the entire spatial profile (Fig. 2d). Because transit delays in conventional drift- or cold-diffusion-limited models scale linearly or quadratically with distance [5], this spatial invariance of the relaxation lifetime demonstrates that carrier extraction does not involve channel-length-dependent transit.

Bias-dependent modulation measurements provide decisive evidence for the non-equilibrium nature of the extraction (Fig. 3a). Bi-exponential fitting deconvolves the transient current into distinct fast ($I_{fast}$) and slow ($I_{slow}$) channels (Fig. 3b). The slow component scales linearly with the applied bias and reverses polarity under opposite voltages, consistent with conventional electric-field-driven photoconductive (PC) drift. In stark contrast, the fast component exhibits complete bias immunity: its polarity, amplitude, and relaxation lifetime remain virtually unchanged regardless of the sign or magnitude of the applied bias (Fig. 3c-f).

These observations—barrier suppression, spatial lifetime invariance, and bias immunity—demonstrate that charge transfer is driven by hot-carrier super-diffusion [14]. Photoexcitation generates an energetic, non-thermalized carrier distribution. Before dissipating their excess energy into the crystal lattice, these hot carriers possess a diffusion coefficient order of magnitude higher than that of cold band-edge carriers ($D_{hot} \gg D_{cold}$) [14, 21]. Driven by the steep electronic thermodynamic gradient ($\nabla Te$) at the illuminated interface, these energetic carriers diffuse super-diffusively into the metal contact prior to energy relaxation, enabling multi-GHz bandwidths in low-mobility TMDCs [14].

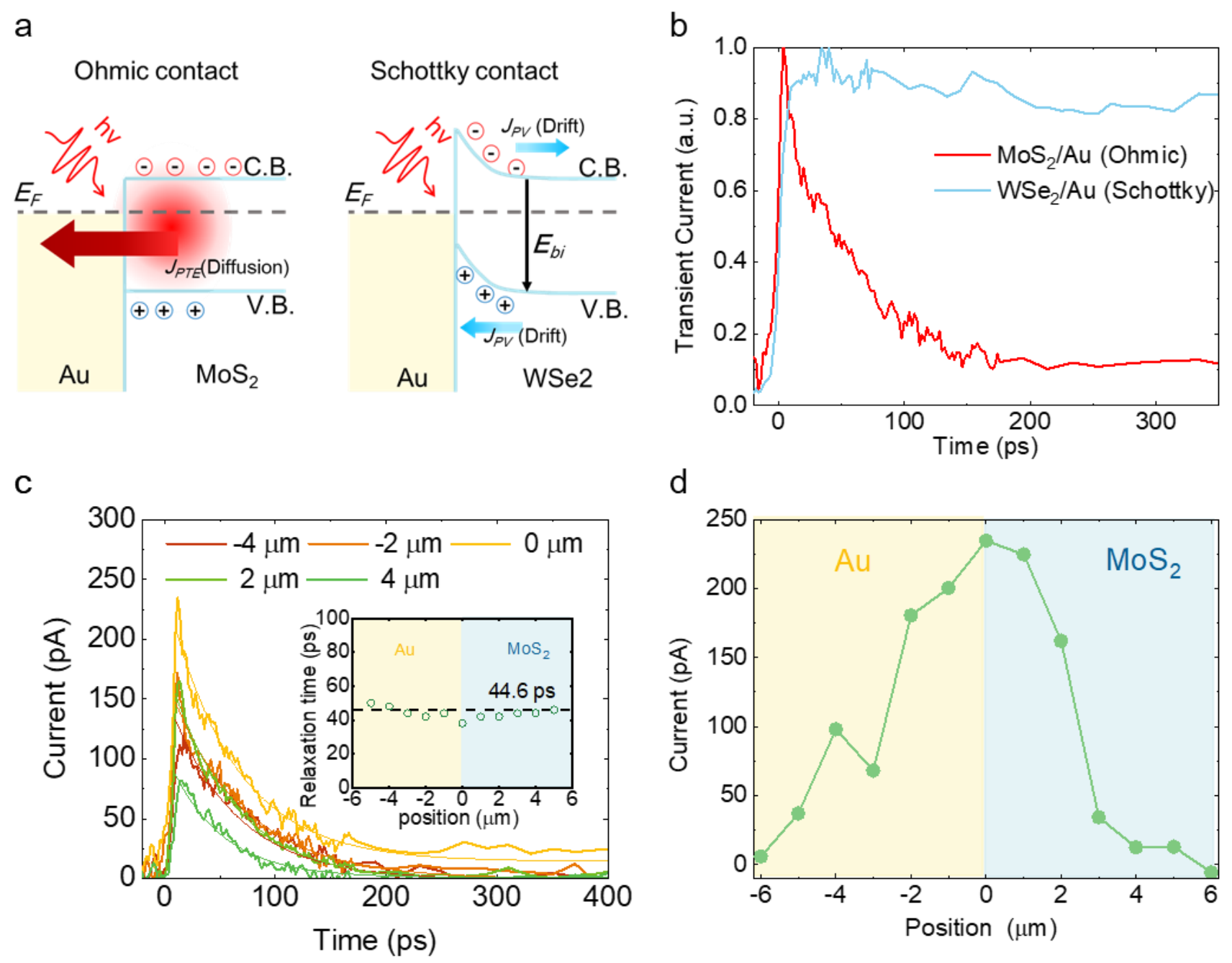

**Figure 2 | Mechanistic and spatial origins of the ultrafast interfacial photocurrent.**

**a.** Schematic illustration of carrier transport mechanisms: hot-carrier super-diffusion under an Ohmic contact (exemplified by $MoS_2$/Au) and photovoltaic (PV) drift under a Schottky contact (exemplified by $WSe_2$/Au). **b.** Normalized time-resolved photocurrents for the $MoS_2$/Au and $WSe_2$/Au devices, with optical excitation focused precisely at the metal-semiconductor interface. **c.** Position-dependent time-resolved photocurrent traces scanned across the device. The $MoS_2$/Au interface is defined as the spatial origin (0 μm), with positive and negative coordinates corresponding to illumination on the $MoS_2$ channel and the Au electrode, respectively. Inset: extracted decay lifetime plotted as a function of illumination position. **d.** Maximum photocurrent intensity plotted as a function of spatial illumination position.

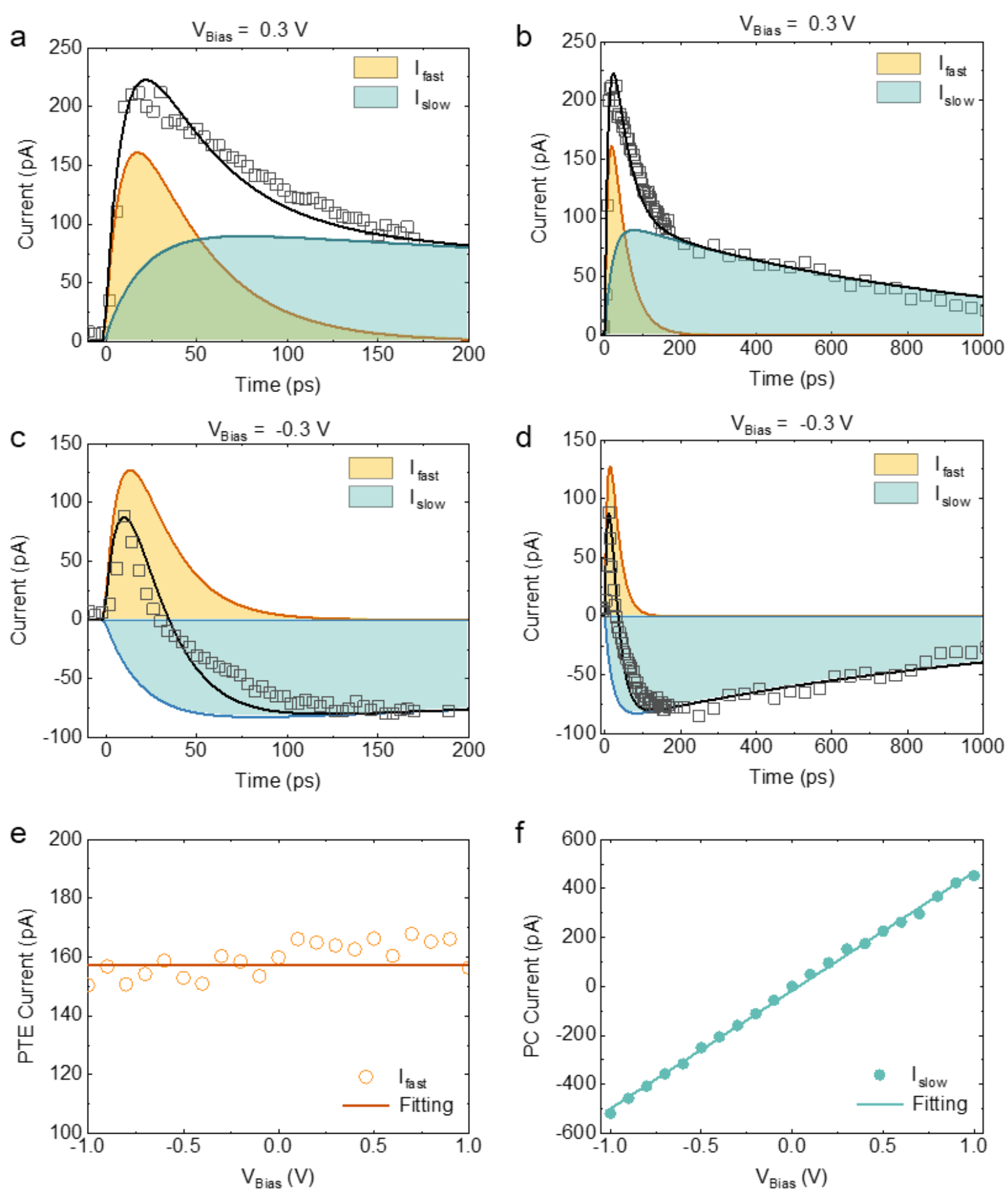

**Figure 3 | Decoupling the photothermoelectric and photoconductive effects via time-resolved photocurrents.**

**a**. Time-resolved transient photocurrent trace of the $MoS_2$/Au device, measured at $V_{Bias}$ = 0.3 V. **b.** The same measurement as in **a**, displayed over an extended timescale of 1,000 ps. **c. d**. Time-resolved transient photocurrent traces measured at $V_{Bias}$ = - 0.3 V on the short (**c**) and extended (**d**) timescales. The solid black lines represent the overall fits to the experimental data, which are decoupled into a fast component ($I_{fast}$, shaded in yellow) and a slow component ($I_{slow}$, shaded in cyan). **e**. Extracted response time of the fast component ($I_{fast}$) as a function of the applied bias voltage. **f.** Extracted photocurrent amplitude of the slow component ($I_{slow}$) as a function of the applied bias voltage.

*Phonon Bottleneck Regulating Super-Diffusion Dynamics*

While super-diffusion accounts for circumventing low band-edge mobility, it raises a fundamental physical question: in typical low-dimensional systems, hot carriers rapidly cool within hundreds of femtoseconds; what sustains this non-equilibrium super-diffusion window in $MoS_2$ for tens of picoseconds (48.5 ps)? The duration of super-diffusion is fundamentally governed by the timescale over which carriers remain in an energetic, non-equilibrium state [14, 21]. In polar layered $MoS_2$, the substantial atomic mass disparity between heavy molybdenum (Mo) and light sulfur (S) atoms opens an acoustic-optical phonon bandgap of ~1.0 THz in the phonon dispersion spectrum (Fig. 4a) [5,6]. This energy gap severely impedes the Klemens decay channel, wherein zone-center optical phonons decay anharmonically into pairs of lower-frequency acoustic modes [23]. As a consequence, non-equilibrium optical phonons emitted during rapid initial electron-phonon thermalization [25] cannot efficiently dissipate into the acoustic lattice bath, triggering a pronounced hot-phonon bottleneck [24,27]. These bottlenecked optical phonons undergo continual re-absorption by non-equilibrium carriers, acting as an internal microscopic energy reservoir that sustains hot-carrier super-diffusion over picosecond timescales (Fig. 4b) [14, 27].

To establish the quantitative link between microscopic lattice dynamics and macroscopic electrical extraction, we performed transient differential reflectance (ΔR/R0) pump-probe spectroscopy across $MoS_2$ flakes of varying thicknesses (5 nm, 13 nm, and 30 nm) and compared the optical relaxation kinetics directly with the on-chip photocurrent decay profiles (Fig. 4c) [26]. As the flake thickness increases, the microscopic optical-to-acoustic phonon scattering lifetime ($\tau_{ph\text{-}ph}$) systematically varies; remarkably, the macroscopic electrical extraction lifetime (τextraction) quantitatively tracks $\tau_{ph\text{-}ph}$ point-by-point across all sample thicknesses (Fig. 4d). This quantitative synchronization demonstrates that the dissipation rate of the hot-phonon bottleneck constitutes the rate-limiting step for interfacial charge extraction ($\tau_{extraction} \approx \tau_{ph\text{-}ph}$).

Comparative measurements on $MoSe_2$ corroborate this phonon-governed transport framework. In $MoSe_2$, the heavier selenium (Se) atoms reduce the cation-anion mass disparity, narrowing the acoustic-optical phonon bandgap [5,6]. This accelerates anharmonic phonon decay and weakens the hot-phonon bottleneck, leading to significantly faster hot-carrier cooling [28]. With accelerated cooling, the hot-carrier extraction lifetime in the $MoSe_2$/Ag device is compressed to 14.2 ps, correspondingly expanding the intrinsic 3-dB bandwidth to 7.5 GHz.

Together, these cross-thickness and cross-material findings establish "phonon engineering" as an optoelectronic design paradigm. Rather than depending solely on electrostatic gating or chemical doping, the operational speed of 2D devices can be engineered by tailoring lattice vibrational spectra, atomic mass disparities, and phonon bandgaps, providing an orthogonal physical approach for zero-bias, ultrafast optoelectronics.

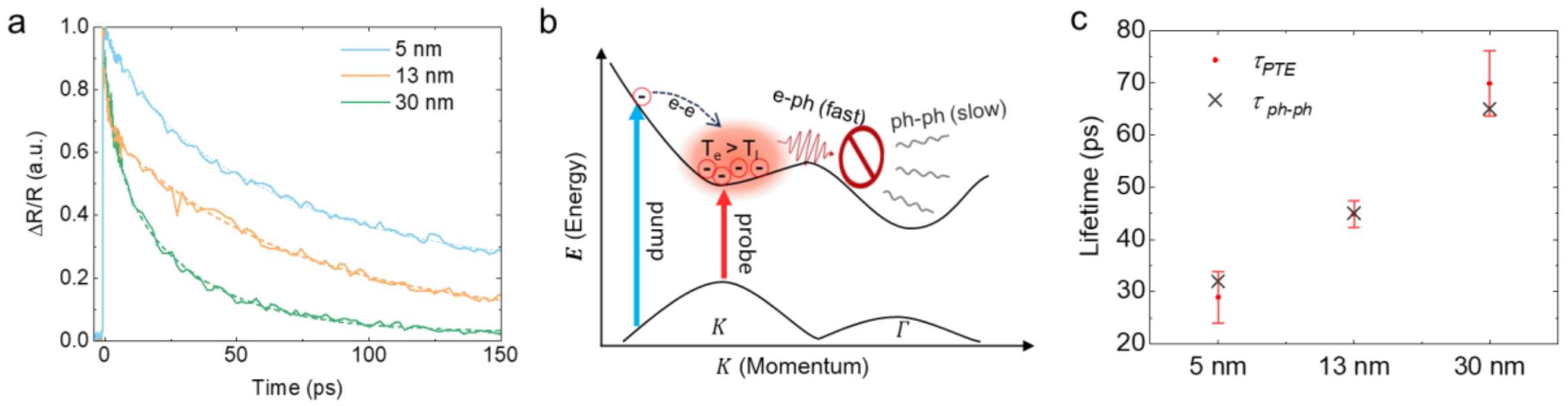

**Figure 4 | Phonon bottleneck governing ultrafast photothermoelectric dynamics.**

**a**. Normalized transient differential reflectance kinetics for $MoS_2$/Au devices with varying $MoS_2$ thicknesses. The pump and probe wavelengths are 400 nm and 680 nm, respectively. **b**. Schematic illustration of the carrier relaxation mechanism in $MoS_2$ following photoexcitation. **c**. Quantitative correlation between the optical-to-acoustic phonon scattering lifetime and the macroscopic PTE relaxation time. Error bars for the PTE response times are extracted from exponential fits.

## Discussion

By combining on-chip terahertz optoelectronic sampling with thermally evaporated Ohmic contacts, we have resolved the intrinsic ultrafast carrier extraction dynamics in unencapsulated TMDCs, circumventing the external parasitic RC delays and transit-time constraints inherent to conventional electric-field-driven drift [12,17,18]. Quantitative cross-validation between bias-dependent electrical sampling and femtosecond pump-probe spectroscopy confirms that this bias-immune, position-independent photoresponse is governed by hot-carrier super-diffusion driven by an interfacial electronic temperature gradient [11,14,21]. Critically, the operational timescale of this ultrafast optoelectronic conversion is tethered to the microscopic optical-to-acoustic phonon scattering lifetime, dictated by the hot-phonon bottleneck that originates from the atomic mass disparity in polar layered lattices [23,24,27]. This correlation establishes a direct physical mapping between the macroscopic operational bandwidth of a 2D optoelectronic device and microscopic anharmonic phonon decay.

From an engineering perspective, phonon-bottleneck-regulated non-equilibrium transport circumvents the speed ceilings imposed by low room-temperature band-edge mobilities, delivering multi-GHz intrinsic bandwidths under zero bias while simultaneously satisfying demands for ultrafast speed, vanishing dark current, and self-powered operation [1,7,12]. More fundamentally, these insights establish "phonon engineering" as a design paradigm for high-speed low-dimensional optoelectronics. Rather than relying on electrostatic gating or chemical doping to boost carrier mobility, the response timescales of optoelectronic conversion can be deterministically tailored by manipulating lattice vibrational spectra, atomic mass disparities, and acoustic-optical phonon bandgaps [5,6,24]. This provides a viable blueprint for breaking traditional transport bottlenecks and realizing next-generation ultra-low-power, chip-scale ultrafast optoelectronic systems [29,30].

## Methods

*Device fabrication*

High-quality flakes of $MoS_2$ and $MoSe_2$ were mechanically exfoliated onto degenerate Si substrates covered with 285-nm $SiO_2$, with flake thicknesses and surface morphologies confirmed via optical microscopy and atomic force microscopy (AFM). Source and drain contact patterns were defined by electron-beam lithography (EBL). Contact electrodes were subsequently deposited via vacuum thermal evaporation at a base pressure below $1 \times 10^{-4}$ Pa: Ni/Au (10 nm/200 nm) for $MoS_2$ channels and Ag (150 nm) for $MoSe_2$ channels. The deposition rates were maintained at 0.05 nm/s for Ni and 0.03 nm/s for Au and Ag to establish intimate metal-semiconductor interfaces and tune the pinning barrier, yielding low-resistance Ohmic contacts. Following lift-off in acetone, the fabricated devices were rinsed and dried with high-purity nitrogen gas.

*On-chip terahertz spectroscopy (electrical readout)*

Intrinsic ultrafast optical-to-electrical conversion dynamics were directly interrogated using on-chip integrated low-temperature-grown GaAs (LT-GaAs) photoconductive switches. The sampling chips were prepared by molecular beam epitaxy (MBE) growth of a 2-μm-thick LT-GaAs film on semi-insulating GaAs substrates at ~300 °C, exhibiting a defect-assisted non-equilibrium carrier lifetime of <2 ps. A femtosecond laser operating at a center wavelength of 680 nm (pulse duration ~100 fs, repetition rate 80 MHz) served as the synchronized optical source. The laser output was split by a polarizing beam splitter (PBS) into pump and probe paths: the pump beam was modulated by an optical chopper at 1 kHz and focused onto the metal-semiconductor edge using a 50× objective lens (NA = 0.5, average power 1.0 mW) to excite transient photocurrents; the probe beam was directed through a motorized optical delay line (Thorlabs DDS220) to trigger the LT-GaAs switch for time-resolved sampling. The sampled transient electrical signals were detected using a dual-phase lock-in amplifier (SRS SR830) under ambient conditions and strict zero bias ($V_{Bias}$ = 0 V).

*Ultrafast Pump-Probe Spectroscopy Measurements*

Transient carrier and lattice cooling dynamics were characterized by femtosecond transient differential reflectance (ΔR/R0) spectroscopy driven by a mode-locked Ti:sapphire oscillator (800 nm, 80 MHz, 150 fs). The primary beam was frequency-doubled via a BBO crystal to generate a 400 nm pump beam, while the secondary beam pumped an optical parametric oscillator (OPO) followed by frequency doubling to deliver a 680 nm probe beam resonant with the A-exciton transition. Both beams were coaxially focused onto the device channel using a 50× objective lens with the pump fluence set at 10 times that of the probe. The pump beam was chopped at 1 kHz (SRS SR540) as the reference for a lock-in amplifier (SRS SR830), and the pump-induced reflectance changes were recorded using a balanced amplified photodetector (Thorlabs PDB210A) paired with a motorized optical delay line, with the temporal decay profiles fitted using a bi-exponential relaxation function.

**Author contributions**:

Z. H. N., J. P. L., and D. Y. W. supervised the project. Y. W. Z., D. Y. W., J. P. L., and Z. H. N. conceived the core ideas, proposed the novel concepts, and designed the overall framework. Y. W. Z. coordinated and conducted the overall experimental investigations. Y. W. Z. and T. Z. carried out the optical spectroscopy measurements. Y. W. Z. and H. W. performed device fabrication and electrical optoelectronic sampling experiments. Y. W. Z. and D. Y. W. discussed the data and drafted the manuscript. All authors discussed the results and have given approval to the final version of the manuscript.

**Competing interests**: The authors declare no competing financial interests.

**Data Availability**: The data that support the findings of this study are available from the corresponding author upon reasonable request.